# Ermittlung und Kommunikation von Anforderungen in etablierten UX-Prozessen

Hartmut Schmitt[1], Gerald Heller[2], Anne Hess[3], Oliver Karras[4]
[1]HK Business Solutions GmbH, hartmut.schmitt@hk-bs.de
[2]Berater und Trainer, gerald.heller@swq4all.de
[3]Fraunhofer IESE, anne.hess@iese.fraunhofer.de
[4] TIB – Leibniz Information Centre for Science and Technology, oliver.karras@tib.eu

Zwischen Requirements Engineering (RE) und User Experience (UX) gibt es starke inhaltliche Überschneidungen. Dennoch werden beide Disziplinen in der Praxis oft von getrennten Rollen ausgeübt und es gibt Defizite in der Zusammenarbeit. Um Anknüpfungspunkte zur Weiterentwicklung von Rollen, Aktivitäten und Artefakten der Disziplinen zu liefern, führt der Arbeitskreis „Requirements Engineering und User Experience" (AK REUX) seit 2021 eine Reihe von Fallstudien durch, in denen die UX-Prozesse verschiedener Unternehmen aus RE-Sicht analysiert werden. Zwischenergebnisse dieser Untersuchung haben wir beim RE-Fachgruppentreffen 2022 vorgestellt und mit den Erfahrungen der Teilnehmer*innen abgeglichen.

**Kontext und Motivation**

Die Wahrnehmungen und Reaktionen der Benutzer*innen, die sich aus der (erwarteten) Benutzung eines interaktiven Systems ergeben – in ihrer Gesamtheit als User Experience bezeichnet [1] – korrespondieren stark mit den Anforderungen der Benutzer*innen. Zur Ermittlung dieser Anforderungen werden im UX teils die gleichen Befragungs-, Beobachtungs- und Kreativitätstechniken genutzt wie im RE. Dennoch fehlt trotz dieser Überschneidungen oft ein gemeinsames Verständnis der Disziplinen, sodass mögliche Synergien verpuffen.

Um dies näher zu beleuchten, führte der AK REUX 2019 eine Onlinestudie unter UX-Professionals durch. Ziel war es, das Arbeitsfeld von UX-Professionals zu beschreiben – insbesondere, wie die Professionals mit ihrem beruflichen Umfeld und ihren Kolleg*innen interagieren – und Einblicke in die konkrete Projektarbeit im Bereich Software- und Systementwicklung zu geben. Insgesamt beteiligten sich 123 UX-Professionals; die Ergebnisse wurden 2019 auf der „Mensch und Computer" [2] und dem RE-Fachgruppentreffen [3] vorgestellt.

Die Ergebnisse dieser Onlinestudie zeigten, dass sich durch eine intensivere Zusammenarbeit zwischen RE und UX erhebliches Optimierungspotential ergeben sollte. Beispielsweise rangiert der Requirements Engineer bei den Rollen, die Anforderungen liefern, auf dem letzten Platz: Lediglich 4 % der befragten UX-Professionals erhalten ihre Anforderungen meist von dieser Rolle, 54 % dagegen nie. Mögliche Erklärungsversuche – etwa, dass es in den Unternehmen der Befragten keine dedizierte RE-Rolle gibt – konnten wir anhand des vorliegenden Datenmaterials allerdings nicht validieren. Daher haben wir uns entschlossen, basierend auf den Ergebnissen der Onlinestudie eine qualitative Studie durchzuführen, in der wir die UX-Prozesse verschiedener Unternehmen detaillierter analysieren.

**Aufbau und Durchführung der Fallstudien**

Insgesamt konnten wir für unsere Fallstudien fünf Teilnehmer*innen gewinnen: je eine Person aus einem deutschen Finanztechnologie-Unternehmen und einem international tätigen Softwareunternehmen (beides Organisationen mit etablierten In-house-UX-Prozessen) sowie drei Personen aus einer deutschlandweit tätigen UX-Agentur. Nachdem wir die Zusage der Teilnehmer*innen erhalten hatten, führten wir zunächst eine *Vorabbefragung* per E-Mail durch. Diese umfasste insgesamt 17 Fragen zum Unternehmen (Größe, Branche, Produkte), zu dessen Organisation und Entwicklungsprozess, zu den genutzten RE-Techniken und -Artefakten sowie zur eigenen Person (Rolle, Aktivitäten, Verantwortlichkeiten) und zum eigenen Team (Größe, regionale Verteilung).

Für die halbstrukturierten Interviews hatten wir einen *Interviewleitfaden* vorbereitet. Dieser Leitfaden umfasste sechs größere Themenblöcke, in denen die Befragten jeweils schildern konnten, was gut bzw. was schlecht funktioniert und welche konkreten Verbesserungspotentiale sie sehen. Zunächst sollten die Befragten ein für ihre Arbeit „typisches Projekt" beschreiben. Daran schlossen sich Fragen zu den typischen „Projektteams und Rollen" in diesen Projekten an. Der nächste Abschnitt widmete sich dem Thema „Anforderungserhebung", insbesondere den hierbei genutzten Anforderungsquellen und Erhebungstechniken, gefolgt von der „Anforderungsdokumentation" mit Fragen zu verwendeten Artefakten und zur Anforderungsqualität. Darauf folgte ein Abschnitt mit Fragen zur „Kommunikation von Anforderungen", z. B. zu den genutzten Kanälen und zur räumlichen Nähe bzw.

Distanz der Akteure. Der letzte Fragenblock widmete sich, da viele Unternehmen mittlerweile agile Vorgehensmodelle adaptiert haben, dem Thema „RE und UX im agilen Umfeld".

Vier Interviews wurden von Zweierteams durchgeführt, wobei eine Person die *Befragung durchführte* und eine Person *protokollierte*. Alle Teilnehmer*innen wurden einzeln befragt. Aufgrund der Corona-Beschränkungen nutzten wir für die Durchführung der Interviews ein Videokonferenz-System. Bei zwei Interviews wurde der *Ton aufgezeichnet*, wodurch das spätere Klären von Unklarheiten vereinfacht wurde; bei zwei Interviews wurde auf eine Aufzeichnung verzichtet, damit die Teilnehmer*innen möglichst unbefangen antworten. Ein Interview konnte vor Ort beim Teilnehmer durchgeführt werden; dieses Interview wurde von einer Einzelperson geführt und ebenfalls aufgezeichnet. Qualitative Unterschiede der Interviews durch die unterschiedlichen Rahmenbedingungen bei den Interviews gab es nicht. Nachdem die Interviews vollständig verschriftlicht waren, wurden die *Ergebnisse validiert* und bei Bedarf mit den Befragten *offene Punkte geklärt*.

Im nächsten Schritt haben wir die *Interviewaussagen vereinzelt* und in Form von Sticky Notes in einem Online-Whiteboard angebracht. Zudem haben wir die Aussagen mit Tags *klassifiziert* (z. B. Aktivitäten, Artefakte) bzw. *bewertet* (z. B. Herausforderungen, interessante Aspekte) , siehe Abbildung 1.

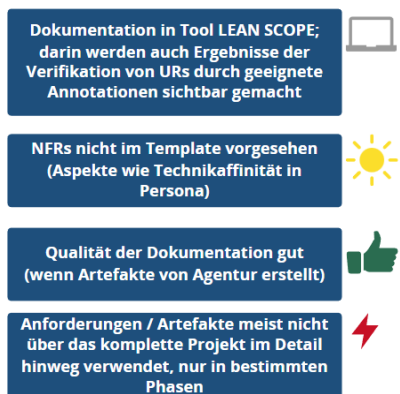

Abbildung 1: Interviewaussagen mit Tags (Beispiele)

Zuletzt haben wir die Informationsflüsse der untersuchten Unternehmen *modelliert*. Hierfür haben wir die *FLOW-Methode* [4] genutzt, die zur Analyse und Verbesserung von Informationsflüssen in Softwareprojekten, insbesondere bei der Kommunikation von Anforderungen, entwickelt wurde. FLOW hat die Besonderheit, dass nicht nur dokumentierte („feste"), sondern auch mündlich weitergegebene („flüssige") Informationen in die Betrachtung einbezogen werden. Dadurch ist FLOW aus unserer Sicht gut zur Dokumentation der Kommunikationsflüsse in den Fallstudien geeignet, vergleiche Abbildung 2.

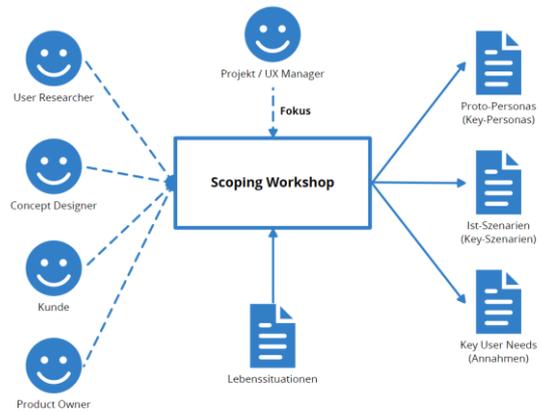

Abbildung 2: Informationsfluss in FLOW (Beispiel)

**Präsentation beim RE-Fachgruppentreffen 2022**

Die bisher erreichten Ergebnisse aus den drei Fallstudien haben wir beim RE-Fachgruppentreffen 2022 vorgestellt [5]. Die Teilnehmer*innen erhielten dadurch Einblick in mehrere etablierte UX-Prozesse und lernten die Techniken kennen, die zur Analyse und Aufbereitung der Ergebnisse genutzt wurden, insbesondere das verwendete Online-Whiteboard und die FLOW-Methode. Der Arbeitskreis erhielt durch die Präsentation Feedback zu den Analyse- und Dokumentationstechniken und zu den Untersuchungsergebnissen, das in die weiteren AK-Arbeiten einfließen wird. Zudem konnten die Ergebnisse mit den Erfahrungen der Teilnehmer*innen hinsichtlich der Zusammenarbeit zwischen RE und UX abgeglichen werden. Als Nächstes sollen die verschiedenen Vorgehen verglichen und generalisierbare Empfehlungen abgeleitet werden.

Mögliche Interessent*innen für weitere Fallstudien bzw. zur Mitwirkung im AK REUX laden wir auf diesem Wege herzlich ein, mit uns in Kontakt zu treten [5]. Unser Dank gilt den Teilnehmer*innen der Fallstudien, ohne deren Bereitschaft die Durchführung nicht möglich gewesen wäre. [Teile dieser Arbeit sind im Rahmen des Projekts „D'accord" (16KIS1506K) entstanden.]

**Referenzen**